\documentclass[11pt]{article}

\usepackage{html,amsmath,amssymb,amsfonts,epsfig,cancel,cite,longtable,caption,siunitx,array,color,booktabs,setspace}

\usepackage{dsfont}

\numberwithin{equation}{section}

\newcommand{\be}{\begin{equation}}
\newcommand{\ee}{\end{equation}}
\newcommand{\beq}{\begin{equation}}
\newcommand{\eeq}{\end{equation}}
\newcommand{\ba}{\begin{array}}
\newcommand{\ea}{\end{array}}
\newcommand{\bea}{\begin{eqnarray}}
\newcommand{\eea}{\end{eqnarray}}
\newcommand{\bean}{\begin{eqnarray*}}
\newcommand{\eean}{\end{eqnarray*}}

\def\e{\epsilon}

\def\gsim{ \lower .75ex \hbox{$\sim$} \llap{\raise .27ex \hbox{$>$}} }
\def\lsim{ \lower .75ex \hbox{$\sim$} \llap{\raise .27ex \hbox{$<$}} }
\def\be{\begin{equation}}
\def\ee{\end{equation}}
\def\bea{\begin{eqnarray}}
\def\eea{\end{eqnarray}}

\newcommand{\pt}{\partial}

\newcommand{\IC}{\mathbb{C}}
\newcommand{\IP}{\mathbb{P}}

\newcommand{\IZ}{\mathbb{Z}}
\newcommand{\cO}{{\cal O}}
\newcommand{\cQ}{{\cal Q}}
\newcommand{\cN}{{\cal N}}

\newcommand{\cA}{{\cal A}}
\newcommand{\cB}{{\cal B}}
\newcommand{\cC}{{\cal C}}
\newcommand{\cL}{{\cal L}}
\newcommand{\cV}{{\cal V}}
\newcommand{\cW}{{\cal W}}

\newcommand{\fn}{\footnotesize}

\def\cjn1{{\cA, \cC^*\otimes \wedge^j \cN^*}}
\def\bjn1{{\cA, \cB^*\otimes \wedge^j \cN^*}}
\def\vjn1{{\cA, \cV^*\otimes \wedge^j \cN^*}}
\def\cjn2{{\cA, \cC\otimes \wedge^j \cN^*}}
\def\bjn2{{\cA, \cB\otimes \wedge^j \cN^*}}
\def\vjn2{{\cA, \cV\otimes \wedge^j \cN^*}}

\newcommand{\cicy}[2]{\begin{matrix} #1\end{matrix}\!\left[\begin{matrix}#2 \end{matrix}\right]}

\newcommand{\varstr}[2]{\vrule height #1 depth #2 width0pt}

\newcommand{\cK}{{\cal K}}

\newcommand{\overbar}[1]{\mkern 3.0mu\overline{\mkern-4.0mu#1\mkern-0.5mu}\mkern 1.5mu}

\renewcommand{\footnoterule}{%
  \kern -3pt
  \hrule width 0.4\textwidth height 0.3pt
  \kern 3pt
}

\begin{document}

\title{{\LARGE \bf$~$\\[-7pt]
A Heterotic QCD Axion\\ [9pt]
}}

\vspace{3cm}

\author{
Evgeny I. Buchbinder${}^{1}$,
Andrei Constantin${}^{2}$,
Andre Lukas${}^{3}$
}
\date{}
\maketitle
\thispagestyle{empty}
\begin{center} { ${}^1${\it The University of Western Australia, \\
35 Stirling Highway, Crawley WA 6009, Australia\\[0.3cm]
       ${}^2$Department of Physics and Astronomy, Uppsala University, \\
       SE-751 20, Uppsala, Sweden\\[0.3cm]
       ${}^3$Rudolf Peierls Centre for Theoretical Physics, Oxford University,\\
       1 Keble Road, Oxford, OX1 3NP, U.K.}}\\
\end{center}

\vspace{11pt}
\abstract
\noindent
We show that a KSVZ axion with a decay constant in the phenomenologically allowed range can be obtained in certain $E_8\times E_8$ heterotic string models. These models have an enhanced symmetry locus in the moduli space, and a non-universal, K\"ahler moduli dependent Fayet-Iliopoulos term which vanishes at this locus. Close to this locus the Fayet-Iliopoulos term is small and can lead to an axion decay constant significantly lower than the string scale. In this way, the no-go arguments of Svr\v{c}ek and Witten, which are based on a universal, dilaton-dependent Fayet-Iliopoulos term, can be avoided. The relevant axion originates from phases of bundle moduli which correspond to deformations away from the enhanced symmetry locus. We construct an explicit example, based on a heterotic line bundle standard model, with all the required ingredients.


\vskip 5cm
{\hbox to 7cm{\hrulefill}}
\noindent{\fn evgeny.buchbinder@uwa.edu.au}\\
{\fn andrei.constantin@physics.uu.se}\\
{\fn lukas@physics.ox.ac.uk}
\newpage


%
%
\section{Introduction}

A Peccei--Quinn axion~\cite{Peccei:1977hh} is considered the most economical solution to the strong CP-problem (see~\cite{Kim:1986ax} for a review and the references therein). The current astrophysical and cosmological bounds impose that the QCD axion decay constant should be in the narrow window $10^9-10^{12}$ GeV, a scale which is difficult to realise in string models with the fundamental scale close to the Planck scale. A detailed study of string theory axions was undertaken by Svr\v{c}ek and Witten in Ref.~\cite{Svrcek:2006yi}. In particular, they showed that in certain heterotic string scenarios it is problematic to find axions with decay constants much smaller that the GUT scale.\footnote{Some string models with a low axion decay constant were studied in~Ref.~\cite{Conlon:2006tq} in the context of Type IIB large volume compactifications and in~Ref.~\cite{Dasgupta:2008hb} in the context of warped heterotic  compactifications. See also Ref.~\cite{Honecker:2013mya} for a recent axion construction in Type IIA string theory and the Ref.~\cite{Choi:2014uaa} for a recent review of QCD axions in string theory.}

In the present note, we propose a heterotic string mechanism which evades the no-go arguments of~Ref.~\cite{Svrcek:2006yi} and can lead to a QCD axion with a small decay constant. The key to our solution is the existence of enhanced symmetry loci in the moduli space of heterotic string compactifications.  These correspond to loci where the heterotic bundle splits into a direct sum of sub-bundles. The enhanced symmetry amounts to one or several $U(1)$ factors, which are generically Green-Schwarz anomalous and have super-heavy associated gauge bosons.

Large classes of phenomenologically interesting heterotic compactifications with such loci are known to exist (see, for example, Ref.~\cite{Anderson:2009nt}). In practice, one starts by constructing the compactification at the enhanced symmetry locus, as it is done for heterotic line bundle models \cite{Anderson:2011ns, Anderson:2012yf, Anderson:2013xka, He:2013ofa, Anderson:2014hia}. The full moduli space in which these special loci reside can subsequently be explored \cite{Buchbinder:2013dna, Buchbinder:2014qda, Buchbinder:2014sya}. Such models are phenomenologically interesting, particularly because the $U(1)$ symmetries can severely constrain the low-energy theory. For example, a supersymmetric standard model with a stable proton has recently been constructed relying on the line bundle approach~\cite{Buchbinder:2013dna, Buchbinder:2014qda, Buchbinder:2014sya}.

We will be working within the context of such $E_8\times E_8$ heterotic Calabi-Yau models equipped with a bundle that splits somewhere in the moduli space. The Fayet-Iliopoulos (FI) term associated to the resulting $U(1)$ symmetry is K\"ahler moduli dependent and vanishes at the split locus. In terms of the underlying 10-dimensional theory, this vanishing property can be understood as the zero-slope condition on the vector bundle. Close to the split locus, the FI term is small but non-zero and the D-term equations can be satisfied by cancelling the FI term with a small vacuum expectation value (VEV) of a bundle modulus. Coupling this bundle modulus to an exotic pair of vector-like quarks \cite{Kim:1979if, Shifman:1979if}, leads to an axion which originates from the bundle moduli phase. Its decay constant is proportional to the bundle moduli VEVs and is, hence, set by the size of the FI term. In this way, an axion with a small decay constant can be obtained close to the split locus in moduli space.

Our note is organised as follows. In Section~\ref{sec:axion_pheno} we describe the mechanism in four-dimensional language. We show that an axion which is the phase of a bundle modulus can indeed couple to QCD and that there are no obstructions to having its decay constant within the observational bound. In Section~\ref{sec:10d} we give the details of the ten-dimensional compactifications of the heterotic string that can lead to the proposed set-up. Finally, in Section~\ref{sec:example}, we provide a concrete example, based on a heterotic line bundle standard model.



\section{The four-dimensional picture}\label{sec:axion_pheno}

We start out by fixing the conventions for the axion and its coupling to QCD.
The QCD axion is a periodic scalar field $\phi$ with a global $U(1)$ shift symmetry $\phi\rightarrow \phi + \text{const}$, which is broken by QCD instanton effects. We normalise the field $\phi$ such that its period is $2\pi$. Moreover, since $\phi$ has mass dimension zero, we can introduce a mass parameter $f$ and define the field $a=f \phi$, which is canonically normalised. Its kinetic term and coupling to QCD are given by the action
\be
S[a]~=\, -\frac{1}{2} \int d^4 x\, \pt_{\mu} a\, \pt^{\mu} a + \frac{r}{8 \pi^2} \frac{1}{f} \int a \  {\rm tr} (F \wedge F)_{\text{\text{QCD}}}\,,
\label{eq:Saxion}
\ee
where $r$ is an integer. We use a normalisation of the field strength $F$ in which the instanton number is given~by
\be
N= \frac{1}{8 \pi^2} \int {\rm tr} (F \wedge F)_{\text{\text{QCD}}}~.
\label{eq:instantonnumber}
\ee

The purpose of this section is to illustrate the mechanism for a heterotic axion with a small decay constant from the point of view of the relevant four-dimensional effective theories. These are $\cN=1$ supergravity theories with the standard model gauge group $G_{\text{SM}}= SU(3) \times SU(2) \times U(1)$ and with one or several additional $U(1)$ symmetries. The additional $U(1)$ symmetries are, generically, anomalous in the Green-Schwarz sense; consequently the associated gauge bosons are massive and, at low-energies, the $U(1)$ symmetries only survive as global symmetries.

Let us now describe the general structure of the spectrum for these theories, focusing on the fields that are relevant to our discussion.
The gravitational spectrum of the model consists of the dilaton, $S= s+i \sqrt{2} \sigma$ (where $\sigma$ is the dilatonic axion), a number of K\"ahler moduli $T^i=t^i+2i\chi^i$ (where $t^i$ are the geometrical fields, measuring the size of Calabi-Yau two-cycles, and $\chi^i$ are the associated axions) plus complex structure moduli which will not play an essential role in our discussion. We assume a general situation with several $U(1)$ symmetries labelled by the index $a$ and with associated gauge fields $A_\mu^a$. Under a gauge transformation $\delta A_\mu^a = -\partial_\mu \eta^a$, the axions $\chi^i$ transform non-linearly
\begin{equation}
 \delta\chi^i=-\epsilon\,{k}_a^i\,\eta^a\; , \label{chishifts}
\end{equation}
where $\epsilon$ is a constant defined in terms of eleven-dimensional quantities, and $k_a^i$ are topological integers defined by the compactification data, as will be discussed in the next section. The dilatonic axion also receives a non-trivial gauge transformation at one-loop level which leads to a one-loop correction to the FI term \cite{Blumenhagen:2005ga}. This correction will not change any of our conclusions, and will henceforth be ignored.

The matter spectrum of the model contains the MSSM fields. In addition, we assume that we have an exotic vector-like pair of quarks, $\cal{Q}-\widetilde{\cal{Q}}$, in order to facilitate the KSVZ mechanism, and several singlet matter fields which correspond to bundle moduli and are neutral under the standard model group. All matter fields~$C^I$, including the aforementioned singlet matter fields, carry charges $q_{a,I}$ under the $a^{\rm th}$ $U(1)$ symmetry and transform linearly as
\beq
\delta C^I = -i\,\eta^a\,q_{a,I}\,C^I\;.
\eeq
The K\"ahler potential for the model is given by
\begin{equation}
 \cK\,=\,-M_{\text{P}}^2\left(\log(S\,+\,\overbar{S})\,+\,\log(\kappa)\,-\,\cK_{\rm cs}\right)\,+\,G_{IJ}\,C^I\,\overbar{C}^J\; ,
 \end{equation}
where $\cK_{\rm cs}$ is the complex structure K\"ahler potential and $C^I$ collectively denote all matter fields listed previously. The specific form of the matter field K\"ahler metric $G_{IJ}$ is not relevant to our discussion and it will be sufficient to know that it is positive definite. The pre-potential, $\kappa$, for the K\"ahler moduli is explicitly given by $\kappa=d_{ijk}\,t^i\,t^j\,t^k$, where the topological numbers $d_{ijk}$ are defined by the underlying string compactification and $\kappa$ is related to the Calabi-Yau volume through the relation $\kappa = \cV/6$. To simplify our discussion, we assume that the K\"ahler moduli space is given by $t^i>0$, which can indeed be achieved in many cases.

From this K\"ahler potential and the $U(1)$ transformations given above, standard four-dimensional supergravity fixes the D-terms, which take the general form~\cite{Anderson:2009nt}
\begin{equation}\label{eq:Dterm}
 D_a=\frac{M^2_P}{\cal V}\, \epsilon\,\,d_{ijk}\,{k_a}^i\, t^j\,t^k - \sum_{I,J}q_{a,I}\,G_{IJ}\,C^I\,\overbar{C}^J~.
\end{equation}

In general, the superpotential $W$ is constrained by the $U(1)$ symmetries. We assume the $U(1)$ charges are such that a cubic coupling between the exotic vector-like quark pair and one of the singlet fields $C$ is allowed. Hence, the superpotential has the form
\beq\label{eq:W}
W =  \lambda{\cal Q}\, C\, \widetilde{\cal Q} + W_{\text{sing}} + \ldots ~,
\eeq
where $W_{\text{sing}}$ is the superpotential for the singlet matter fields and the dots refer to the usual MSSM superpotential terms. The coupling ${\cal Q}\, C\, \widetilde{\cal Q}$ will be crucial in our discussion of the axion mechanism. In addition, we assume that the singlet superpotential is such that the field $C$ remains an F-flat direction.

The final ingredient of the low-energy field theory is the gauge kinetic function for the standard model gauge fields, which is universal and is given by
\beq\label{eq:gkf}
f_{\rm SM} = S + \pi\e_S \beta_i\, T^i~.
\eeq
Here $\e_S$ is the strong coupling expansion parameter and $\beta_i$ are topological numbers defined by the compactification data. In the next sections we will show that low-energy theories with all the above ingredients can indeed be obtained from the $E_8\times E_8$ heterotic string and an explicit example will be provided in Section~\ref{sec:example}.

The theory schematically described above contains several axionic fields $\sigma, \chi^i$ which couple to $\text{tr}(F\wedge F)_{\text{QCD}}$ via the gauge kinetic function~\eqref{eq:gkf}. These axions represent the traditional candidates for resolving the strong CP problem within the heterotic string context. However, it has been known for a long time (see, for example, Ref.~\cite{Svrcek:2006yi} and the references therein) that the decay constants for these axions are of order of the GUT scale and, therefore, are too large to comply with the phenomenological constraints.
\vspace{8pt}

This situation is radically different if one considers the axions which are the phases of the singlet fields. We now turn to the discussion of these axions. We write the singlet field as $C= h e^{i\phi}$, so that the $U(1)$ symmetries act on $\phi$ by shifts
\beq
\phi \rightarrow \phi - q_a\, \eta^a\;,
\eeq
where $q_a$ denote the $U(1)$ charges of $C$. It is convenient to define a new basis for the $U(1)$ generators, such that the field $C$ is charged under a single $U(1)$, with a charge that we denote by $q$. Integrating out ${\cal Q}$ and $\widetilde{\cal Q}$ at energies
below $\langle |C|\rangle =h$ gives a contribution to the effective action consistent with the chiral anomaly
\be
-\frac{1}{8 \pi^2} \int \phi \ {\rm tr} (F \wedge F)_{\text{QCD}}\,,
\label{3.7}
\ee
which provides the coupling of the axion $\phi$ to $(F \wedge F)_{\text{QCD}}$. Thus, the effective action for the axion becomes
\be
S[\phi]~=~ -\int d^4 x\, \left( h^2 (\pt_{\mu} \phi  - q A_{\mu})^2 + D^2\right)-\frac{1}{8 \pi^2} \int \phi \ {\rm tr} (F \wedge F)_{\text{QCD}}\,.
\label{3.6}
\ee

The same effect can also be understood from a different viewpoint.
Integrating out a massive ${\cal Q}-\widetilde{\cal Q}$ pair produces a 1-loop  threshold correction to the gauge coupling given by~\cite{Kaplunovsky:1994fg}
\be
- \frac{T (r)}{8\pi^2} \log h  = -\frac{1}{16 \pi^2} \log h \,,
\label{3.8}
\ee
where $T(r)$ is the quadratic Casimir which is equal to $1/2$ if the ${\cal Q}-\widetilde{\cal Q}$ pair transforms in the fundamental (antifundamental) representation of $SU(3)$. This means that we have a contribution to the effective
action of the form
\be
-\frac{1}{16 \pi^2}  \int d^4 x \log h \ {\rm tr} F^2\,.
\label{3.9}
\ee
By supersymmetry this implies~\eqref{3.7}. From Eq.~\eqref{3.6}, we see that $\phi$ indeed couples to QCD and, comparing with Eq.~\eqref{eq:Saxion}, we obtain the axion decay constant
\be
f= \sqrt{2} h\,.
\label{3.10}
\ee
The value of $h$ in the supersymmetric vacuum is controlled by the $D$-term equation (see Eq.~\eqref{eq:Dterm})
\be
D=\frac{M^2_P}{\cal V}\, \epsilon\,\,d_{ijk}\,{k}^i\, t^j\,t^k - q\, h^2 =0\;,
\label{3.11}
\ee
where $(k^i)$ represents a linear combination of the vectors $(k_a^i)$ that corresponds to the linear combination of the $U(1)$ generators discussed above. The first term represents the FI contribution. Provided the vector $(k^i)$ contains both positive and negative entries this FI term  can vanish at a certain locus in moduli space, which we will also refer to as the split locus. This indeed happens for many examples. From a 10-dimensional point of view,  the vanishing of the FI term is linked to the zero-slope condition on the vector bundle, as will be discussed in the next section. It is clear that, at the split locus,  $h$ must vanish in order to preserve supersymmetry.

Moving away from the split locus, the magnitude of the FI term can be smoothly varied in an interval around zero.
Hence, there is no obstruction to having  $h$ small, so that the axion decay constant $f$ is consistent with the observational bound $ 10^9< f<  10^{12}$ GeV.
Note that this mechanism does not work in the case of the universal anomalous $U(1)$ symmetry considered in Ref.~\cite{Svrcek:2006yi}. In this case, the FI term is proportional to $1/s$ and is of the order of the GUT scale as long as the gauge coupling has a value in the phenomenologically required range.
Apart from generating the coupling of $\phi$ to $\text{tr}( F\wedge F)_{\rm QCD}$, the superpotential term ${\cal Q}\,C\widetilde{\cal Q}$ also generates a mass for the exotic vector-like pair well above the TeV scale, thus removing it from the low-energy spectrum.

Moreover, given that the value of $h$ is much below the compactification scale, the mass of the $U(1)$ gauge boson receives its leading contribution from the $\chi^i$ and $\sigma$ kinetic terms, thereby breaking the $U(1)$ gauge symmetry close to the GUT scale. Below this scale, the $U(1)$ appears only as a global symmetry which is then spontaneously broken by a non-vanishing VEV $\langle |C|\rangle =h$.


\section{The higher-dimensional picture}\label{sec:10d}
In this section, we briefly review the structure of $E_8\times E_8$ heterotic string compactifications on Calabi-Yau manifolds with split vector bundles, following Refs.~\cite{Anderson:2011ns,Anderson:2012yf}. Our emphasis will be to show how the various ingredients in the effective four-dimensional theory required for a successful axion model, as described in the previous section, can be obtained in such compactifications.

We consider a compactification of the $E_8\times E_8$ heterotic string (in the weak or strong coupling limit) on a Calabi-Yau (CY) three-fold $X$ with a rank five vector bundle $V\rightarrow X$ which splits as
\begin{equation}
 V=\bigoplus_{a=1}^A V_a
\end{equation}
where $V_a$ are bundles with structure groups $U(n_a)$, subject to the constraints $\sum_a n_a=5$ and $c_1(V)=0$. In this way, the structure group of $V$ is contained in $S(U(n_1)\times\dots\times U(n_A))\subset SU(5)\subset E_8$ whose commutant in $E_8$ -- the observable low-energy gauge group -- is given by $SU(5)\times S(U(1)^A)$. In general, there is another vector bundle in the hidden $E_8$ sector but this will not be relevant to our discussion.

For the above compactification to preserve supersymmetry the bundle $V$ needs to be poly-stable with slope zero. This is equivalent to saying that each sub-bundle $V_a$ is slope-stable and has vanishing slope. The slope of $V_a$ is explicitly given by
\be
\mu (V_a)=\frac{1}{{\rm rk} (V_a)}\int_X c_1 (V_a)\wedge J \wedge J\, = \frac{1}{{\rm rk} (V_a)} d_{ijk}\,c_1^i(V_a)\,t^j\,t^k\stackrel{!}{=}0
\label{2.3}
\ee
where $J=t^iJ_i$ is the K\"ahler form on $X$, the $J_i$, $i=1,\ldots ,h^{1,1}(X)$,  form a basis of the second cohomology of $X$ and $t^i$ are the K\"ahler moduli. We note that the slope is proportional to the numerator of the FI term in Eq.~\eqref{eq:Dterm}, when the topological numbers $k^i_a$ are identified as $k^i_a=c_1^i(V_a)$. In this way, the vanishing of the FI term, which is a crucial ingredient in our scenario, is directly tied to the supersymmetry of the internal vector bundle.

The matter spectrum of the four-dimensional GUT theory with gauge group $SU(5)\times S(U(1)^A)$ is controlled by the cohomology of $V$ and its associated tensor bundles and is summarised in the table below.
\begin{center}
\vspace{8pt}
\begin{tabular}{|c|c|c|c|}\hline
\varstr{12.4pt}{7pt}~~~multiplet~~~&$~~S(U(1)^A)$ charge~~~&~~~~~bundle~~~~~&~~~~~~~cohomology~~~~~~~\\\hline\hline
\varstr{12.4pt}{7pt}${\bf 10}_a$&$~~{\bf e}_a$&$V_a$&$H^1(X,V_a)$\\\hline
\varstr{12.4pt}{7pt}$\overline{\bf 10}_a$&$-{\bf e}_a$&$V_a^*$&$H^1(X,V_a^*)$\\\hline
\varstr{12.4pt}{7pt}$\overline{\bf 5}_{a,b}$&$~~{\bf e}_a+{\bf e}_b$&$V_a\otimes V_b$&$H^1(X,V_a\otimes V_b)$\\\hline
\varstr{12.4pt}{7pt}${\bf 5}_{a,b}$&$-{\bf e}_a-{\bf e}_b$&$V_a^*\otimes V_b^*$&$H^1(X,V_a^*\otimes V_b^*)$\\\hline
\varstr{12.4pt}{7pt}${\bf 1}_{a,b}$&$~~{\bf e}_a-{\bf e}_b$&$V_a\otimes V_b^*$&$H^1(X,V_a\otimes V_b^*)$\\\hline
\end{tabular}
\vspace{8pt}
\end{center}
Here, ${\bf e}_a$ denotes the $a^{\rm th}$ standard unit vector in $A$ dimensions, so that, for example, the multiplet ${\bf 10}_a$ carries charge one under the $a^{\rm th}$ $U(1)$ symmetry and is uncharged under the others. Provided that the Calabi-Yau manifold $X$ has a freely-acting symmetry $\Gamma$ (which lifts to the bundle $V$), the above GUT model can be quotioned by $\Gamma$ and a Wilson line can be introduced in order to break the GUT group to $G_{\rm SM}\times S(U(1)^A)$. Then, the GUT multiplets in the above table break up into the usual standard model multiplets. For a model with a phenomenologically viable field content we require that $h^1(X,V)=3|\Gamma|$ (three ${\bf 10}$ multiplets), $h^1(X,V^*)=0$ (no $\overline{\bf 10}$ multiplets), $h^1(X,\wedge^2 V)=3|\Gamma|+n$ and $h^1(X,\wedge^2 V^*)=n$ (three $\overline{\bf 5}$ multiplets plus whatever remains from the additional $n$ vector-like $\overline{\bf 5}\,$--$\,{\bf 5}$ pairs). The vector-like $\overline{\bf 5}\,$--$\,{\bf 5}$ pairs can lead to a pair of Higgs doublets and, depending on the Wilson line choice, also to a vector-like pair of exotic quarks, as required for our axion models. Whether this can be achieved depends on the details of the model, specifically the Wilson line choice, and a concrete example will be given in the next section.

Further, we note that all matter fields in the above table carry charges under the additional $U(1)$ symmetries. This includes the singlet matter fields ${\bf 1}_{a,b}$, which describe deformations away from the split locus. The existence of a trilinear superpotential coupling between a singlet matter field and the exotic quark pair, which is crucial for the axion model (see Eq.~\eqref{eq:W}), depends on the specific charges of the fields in a given model. However, we note that the general structure of charges, as in the above table, is consistent with such a term. In the next section, we will present an example model where this trilinear term is indeed allowed.

Finally, we should explain the higher-dimensional origin of the FI term in Eq.~\eqref{eq:Dterm}. For this, it is sufficient to explain the non-linear transformations of the axion fields $\chi^i$ in Eq.~\eqref{chishifts}. These fields originate from the M-theory three-form $C$ (here we use the strong-coupling version of the theory, but the weak coupling formulation leads to identical results) as
\be
C_{11 a \bar b}= \chi^i (J_i)_{a \bar b}\; ,
\label{2.9}
\ee
and they combine into four-dimensional supermultiplets as $T^i=t^i+2i\chi^i$. It is a general feature of heterotic theories, induced by the Bianchi identity, that three-form $C$ transforms non-trivially under $E_8\times E_8$ gauge transformations~\cite{Horava:1995qa, Horava:1996ma}. For the present compactifications with split bundles, this implies
\be
\delta C_{11 a\bar b}= -\Big( \frac{\kappa_{11}}{4 \pi}\Big)^{2/3} \frac{1}{4 \pi}\, \delta (x^{11})\, {\rm tr} (\eta F_{a \bar b})\,.
\label{2.11}
\ee
where $F$ is the internal field strength of any of the additional $U(1)$ symmetries, $\eta$ is the corresponding four-dimensional transformation parameter and $\kappa_{11}$ is the 11-dimensional Newton constant. Integrating this equation over CY two-cycles ${\cal C}^i$ dual to the basis $J_i$, as well as over the orbifold $S^1/\mathbb{Z}_2$, and taking into account Eq.~\eqref{2.9}, we have
\be
\delta \chi^i= -\frac{\e}{4 \pi}  \int_{{\cal C}^i } {\rm tr} (\eta F)\,.
\label{2.12}
\ee
where $\epsilon=\epsilon_S\epsilon_R^2$ and
\beq
\epsilon_S = \left( \frac{\kappa_{11}}{4 \pi}\right)^{2/3}  \frac{1}{\pi \rho  v^{1/3}}\;,\quad
\epsilon_R=\frac{v^{1/6}}{\pi\rho}
\eeq
are the relevant expansion parameters in the strong coupling limit \cite{Lukas:1997fg, Lukas:1998hk}. Here $v$ and $\pi\rho$ are the reference volumes of the CY manifold and the orbi-circle, respectively, so that the four-dimensional Planck mass is determined by $M_{\text{P}}^2 =\pi \rho\, v/\kappa_{11}^2$. Eq.~\eqref{2.12} immediately implies the non-linear transformation law \eqref{chishifts} for the axions $\chi^i$, identifying $k_a^i=c_1^i(V_a)$, as before.

In summary, we have seen that all the required ingredients for a successful axion model are present in heterotic CY models with split bundles. We obtain additional, Green-Schwarz anomalous $U(1)$ symmetries with associated FI terms which can vanish at specify loci in K\"ahler moduli space. Standard multiplets as well as additional singlet matter fields are charged under these $U(1)$ symmetries and vector-like pairs of exotic quarks with a trilinear superpotential coupling to a singlet matter field can be obtained for suitable model building choices. In the next section, we will provide an explicit example, in the context of heterotic line bundle bundles, which realises all these properties.


\section{An explicit example}\label{sec:example}
The database \cite{lbdatabase} contains a large number of phenomenologically promising $SU(5)$--GUT models, derived from the $E_8\times E_8$ heterotic string compactified on smooth Calabi-Yau manifolds with line bundle sums. These models have the right field content to lead to three families of quarks and leptons after the inclusion of a Wilson line; they also have a number of vector-like $\overline{\bf 5}$--${\bf 5}$ pairs, intended to account for a pair of Higgs doublets. In the model building approach pursued in Refs.~\cite{Anderson:2011ns,Anderson:2012yf}, the Wilson line was chosen to project out the triplets from the $\overline{\bf 5}$--${\bf 5}$ pairs while keeping at least one pair of Higgs doublets -- clearly the simplest and cleanest way to arrive at an MSSM-like spectrum. In the present context, we will slightly modify this approach  in order to implement the KSVZ axion. We will choose a Wilson line which leads to one pair of Higgs triplets, in addition to the pair of Higgs doublets. The database \cite{lbdatabase} can, in principle, be searched systematically for models which allow for such a choice and we expect that a large number of possibilities will emerge in this way. Here, we are merely interested in a proof of existence and we will, therefore, focus on a single example with the right properties.

\subsection{The manifold}
The model in question is defined as a compactification on a smooth Calabi-Yau threefold $X$ realised as an intersection of two hypersurfaces in a product of five $\mathbb C\mathbb P^1$ spaces, as summarised by the following configuration matrix:
\begin{equation}\label{Xconf}
X~=~~
\cicy{\IC\IP^1 \\   \IC\IP^1\\ \IC\IP^1\\ \IC\IP^1\\ \IC\IP^1}
{ ~1 & 1\!\!\!\! & \\
  ~1 & 1\!\!\!\! & \\
  ~1 & 1\!\!\!\! & \\
  ~1 & 1\!\!\!\! & \\
  ~0 & 2\!\!\!\!}_{-80}^{5,45}\
\end{equation}
Manifolds in this class have Euler number $\eta = - 80$, Hodge numbers $h^{1,1}(X)=5$ and $h^{2,1}(X)=45$.  A basis $\{J_i\}$ of the second cohomology is provided by the pull-backs of the hyperplane classes of the five $\IC\IP^1$ spaces. We can expand the K\"ahler forms on $X$ as $J=t^i\,J_i$, where $t^i$ are the K\"ahler moduli whose K\"ahler cone is defined by $t^i\geq 0$. Relative to the basis $\{J_i\}$, the triple intersection numbers have the following simple form
\beq
d_{ijk} = \int_X J_i\wedge J_j\wedge J_k = \begin{cases} ~2 & \mbox{ if } i\neq j, j\neq k \\ ~0 &\mbox{ otherwise } \end{cases}\; .\label{tqisec}
\eeq
The second Chern class of the tangent bundle is given by $c_2(TX)=(24,24,24,24,24)$, relative to a basis of the fourth cohomology dual to $\{J_i\}$. We will denote line bundles $\cL$ with first Chern class $c_1(\cL)=k^iJ_i$ by $\cL={\cal O}_X({\bf k})$. Then, from Eq.~\eqref{Xconf},  $X$ is defined as the common zero set of two sections $p_1\in \Gamma(\cO_X(1,1,1,1,0))$ and $p_2\in \Gamma (\cO_X(1,1,1,1,2))$. For specific choices of these sections, $X$ has a freely-acting $\mathbb{Z}_2\times \mathbb{Z}_2$ symmetry~\cite{Braun:2010vc}. Denoting by $x_{m,0}$, $x_{m,1}$ the homogeneous coordinates of the $m$-th projective space, the action of the two generators on these coordinates is given by
\beq\label{eq:groupaction}
g_1:~ x_{m,\alpha}\mapsto (-1)^\alpha x_{m,\alpha}~, \qquad g_2:~ x_{m,\alpha}\mapsto x_{m,\alpha+1}\; ,
\eeq
where the index $\alpha$ is understood to take values in $\IZ_2$. At the same time, the generators act on the two defining polynomials as $\tilde{g}_1={\rm diag}(1,-1)$ and $\tilde{g}_2={\rm diag}(-1,1)$. The quotient manifold $\widehat{X}=X/(\IZ_2\times \IZ_2)$ has a non-trivial fundamental group, $\pi_1(\widehat{X}) = \IZ_2\times \IZ_2$, and allows for the introduction of discrete Wilson lines. This manifold has Hodge numbers $h^{1,1}(\widehat X)=5$ and $h^{2,1}(\widehat X)=15$, as can be computed following the methods used in Refs.~\cite{Candelas:2008wb, Candelas:2010ve}.

\subsection{The GUT model at the split locus}
The bundle $V$ is chosen as a sum of five line bundles
\beq
V ~=~ \bigoplus_{a=1}^5 \,\cL_a  ~ = ~ \bigoplus_{a=1}^5 \,\cO_X({\bf k}_a)
\eeq
explicitly given by
\begin{equation}\label{eq:kmatrix}
({k^i}_a)=
\cicy{ \\ \\ \\ \\ }
{ -2 & ~~1 & ~~1 & ~~0 & ~~0~ \\
~~1 & -2 & ~~0 & ~~1 & ~~0~ \\
~~0 & ~~1 & -2 & ~~0 & ~~1 ~\\
 ~~1 & ~~0 &~~ 0 & -1 & ~~0 ~\\
 ~~0 & ~~0 &~~ 1 & ~~0 & -1 ~\\}\; .
\end{equation}
Since the columns of this matrix sum up to zero we have $c_1(V)=0$ and the structure group is given by $S\big(U(1)^5\big)\subset SU(5)$. The second Chern class of this bundle is given by $c_2(V)=(10,10,10,18,18)$ and, hence, comparing with $c_2(TX)$, we see that it is consistent with anomaly cancelation. The index of $V$ is $\chi(V)=-12$, appropriate for obtained a three-family model after dividing by the order four symmetry $\mathbb{Z}_2\times \mathbb{Z}_2$. Using the definition~\eqref{2.3} and the explicit values~\eqref{tqisec} for the intersection numbers, it is easy to show that the slopes, $\mu(\cL_a)$, of these five line bundles vanish simultaneously at the split locus $t_1=t_2=t_3=t_4=t_5$. At this locus, the low-energy GUT group is $SU(5)\times S(U(1)^5)$ (with all $U(1)$ vector bosons massive, as can be seen by inspecting the rank of the matrix~\eqref{eq:kmatrix}), and the matter spectrum is given by
\begin{gather}
4\,\mathbf{10}_1, ~4\,\mathbf{10}_2, ~4\,\mathbf{10}_3, \nonumber\\[4pt]
~4\,\overline{\mathbf{5}}_{1,3}, ~4\,\overline{\mathbf{5}}_{1,5}, ~4\,\overline{\mathbf{5}}_{3,4}, ~3\,\mathbf{5}_{1,4}, ~3\,\overline{\mathbf{5}}_{1,4}, ~\mathbf{5}_{4,5}, ~\overline{\mathbf{5}}_{4,5},\\[4pt]
12\,\mathbf{1}_{1,3},  ~4\,\mathbf{1}_{1,5}, ~16\, \mathbf{1}_{2,4}, ~4\,\mathbf{1}_{2,5}, ~12\, \mathbf{1}_{3,2}, ~4\,\mathbf{1}_{3,4}, ~12\, \mathbf{1}_{3,5}, ~3\,\mathbf{1}_{1,4}, ~3\,\mathbf{1}_{4,1},~\mathbf{1}_{4,5},~\mathbf{1}_{5,4}  \nonumber
\end{gather}
Evidently, we have 12 chiral families in ${\bf 10}\oplus\overline{\bf 5}$, which will lead to three families after carrying out the $\mathbb{Z}_2\times\mathbb{Z}_2$ quotient, plus four vector-like $\overline{\bf 5}\,$--$\,{\bf 5}$ pairs and a spectrum of singlet matter fields. It is important that we have $\overline{\bf 5}\,$--$\,{\bf 5}$ pairs from two different $U(1)$ charge sectors, one of which can lead to the Higgs doublets, the other one to Higgs triplets. In this way, it is possible to have a  trilinear superpotential coupling between the Higgs triplets and a singlet matter field but avoid the analogous trilinear coupling between the Higgs doublets and the same singlet matter field.

\subsection{The MSSM with a vector-like pair of quarks}

The line bundle sum $V$ descends to a bundle $\widehat{V}$ on the quotient manifold $\widehat{X}=X/(\IZ_2\times \IZ_2)$ if and only if it has a $(\IZ_2\times \IZ_2)$--equivariant structure. Moreover, when each line bundle $\cL_a$ is individually equivariant, as will be the case for our example, the bundle $\widehat{V}$ is also a direct sum of line bundles. As a result, the number of $U(1)$ symmetries and the $U(1)$ charges of the various multiplets remain unchanged after taking the quotient.

It can be checked that all five line bundles $\cL_a$ in Eq.~\eqref{eq:kmatrix} admit an equivariant structure with respect to the group action \eqref{eq:groupaction}. However, this equivariant structure is not unique, and two equivariant structures can differ by a fiber-wise action of the group. Thus, we can classify the equivariant structures of the line bundles~$\cL_a$ by five irreducible $\mathbb{Z}_2\times\mathbb{Z}_2$ representations. In general we denote $\mathbb{Z}_2\times\mathbb{Z}_2$ representations by $(p,q)$, where $p,q=0,1$ and also introduce the regular representation ${\cal R}$ and the representation $\widetilde{\cal R}$ given by
\be
 {\cal R}=(0,0)\oplus (0,1)\oplus(1,0)\oplus(1,1)\; ,\quad  \widetilde{\cal R}=  (0,0)\oplus (1,0)\oplus (1,1)\; .
\ee
For our specific model, we choose the following equivariant structure:
\beq
\cL_1^{(0,1)}\oplus \cL_2^{(0,0)}\oplus \cL_3^{(0,0)}\oplus \cL_4^{(0,0)}\oplus \cL_5^{(0,0)}~.
\eeq
Given this choice, we can compute the decomposition of the relevant cohomologies into  $\mathbb{Z}_2\times\mathbb{Z}_2$ representations. These are given by
\begin{equation*}
\begin{array}{rrlrrrlrrrl}
 H^1(X,\cL_1)&=&{\cal R}&\quad& H^1(X,\cL_2)&=&{\cal R}&\quad& H^1(X,\cL_3)&=& \cal R\\[4pt]
 H^1(X,\cL_1\otimes \cL_3)&=& \cal R&\quad&H^1(X,\cL_1\otimes \cL_5)&=&{\cal R}&\quad&H^1(X,\cL_3\otimes \cL_4)&=&{\cal R}\\[4pt]
 \end{array}
 \vspace{-12pt}
\end{equation*}
\begin{equation}\label{lbcohgrad}
\begin{array}{rrlrrrl}
 H^1(X,\cL_1\otimes \cL_4)&=&\widetilde{\cal R}&\quad&H^1(X,\cL_1^*\otimes \cL_4^*)&=&\widetilde{\cal R} \\[4pt]
 H^1(X,\cL_4\otimes \cL_5)&=&(0,1)&\quad&H^1(X,\cL_4^*\otimes \cL_5^*)&=&(0,1) \\[4pt]
 H^1(X,\cL_1\otimes \cL_4^*)&=&\widetilde{\cal R}&\quad&H^1(X,\cL_1^*\otimes \cL_4)&=&\widetilde{\cal R} \\[4pt]
 H^1(X,\cL_4\otimes \cL_5^*)&=&(0,1)&\quad&H^1(X,\cL_4^*\otimes \cL_5)&=&(0,1) \\[4pt]
 \end{array}
\end{equation}
All remaining singlet cohomologies that have been omitted in \eqref{lbcohgrad} correspond to multiples of the regular representation $\cal{R}$.

In order to break the GUT group to the gauge group of the Standard Model and to project out the unwanted states, we complete the bundle on the quotient manifold to $\widehat{V}\oplus \cal{W}$, where $\cal{W}$ is a flat rank one bundle (a Wilson line), with structure group $\IZ_2\times\IZ_2$, embedded in the hypercharge direction of $SU(5)$. The Wilson line can be specified by two irreducible $\mathbb{Z}_2\times\mathbb{Z}_2$ representations, denoted by ${\cal W}_2$ and ${\cal W}_3$, satisfying ${\cal W}_2\neq {\cal W}_3$ and ${\cal W}_2^{\otimes 2}\otimes {\cal W}_3^{\otimes 3}=(0,0)$.  We aim to obtain the exact chiral matter spectrum of the MSSM, with the chiral $SU(5)$--multiplets being broken in the usual way as $\overline{\bf 5}_{a,b}\rightarrow (d_{a,b},L_{a,b})$ and ${\bf 10}_a\rightarrow (Q_a,u_a,e_a)$. In addition, we would like to project out the triplets from the $(\overline{\mathbf 5}_{4,5}, {\mathbf 5}_{4,5})$ vector-like pair and retain a pair of Higgs doublets. From the three $(\overline{\mathbf 5}_{1,4}, {\mathbf 5}_{1,4})$ vector-like pairs we would like to retain a single vector-like pair of triplets $T-\overline{T}$ (exotic quarks). The appropriate choice of Wilson line is given by
\beq
{\cal W}_2=(0,1)\;,\quad {\cal W}_3=(0,0)~.
\eeq
The  $\mathbb{Z}_2\times\mathbb{Z}_2$ charges of the various standard model multiplets, including the exotic vector-like quark pair, are listed below:
\begin{equation}\label{Wf}
\begin{array}{rlr}
  \cW(d)=\cW(T)=\overline\cW_3=(0,0)&\quad\quad\quad&\cW(L)=\cW(H)=\overline\cW_2=(0,1)\\[4pt]
  \cW(\overline{ T}) = \cW_3 = (0,0) & \quad & \cW(\overbar H) = \cW_2 = (0,1)\\[4pt]
  \cW(Q)=\cW_2\otimes \cW_3=(0,1) & \quad & \cW(u)=\cW_3\otimes \cW_3=(0,0)\\[4pt]
  &\!\!\!\!\!\!\!\!\!\!\!\!\!\!\!\!\!\!\!\!\!\!\cW(e)=\cW_2\otimes \cW_2=(0,0)~.\!\!\!\!\!\!\!\!\!\!\!\!\!\!\!\!\!\!\!\!\!\!\!\!\!\!\end{array}
\end{equation}

With these charges, we can compute the number of multiplets of any given type $\psi$ resulting from the GUT symmetry breaking. Thus, if $\psi$ is associated with a cohomology group $H^1(X,{\cal L})$, we have to extract the $\mathbb{Z}_2\times\mathbb{Z}_2$ singlets from $H^1(X,{\cal L})\otimes \cW(\psi)$. From Eqs.~\eqref{lbcohgrad}, \eqref{Wf} and the identification of cohomologies and particles discussed in the previous section, we obtain the following standard model spectrum:
 \begin{equation}
 \begin{gathered}
\mathbf{10}_1, ~\mathbf{10}_2, ~\mathbf{10}_3, ~\overline{\mathbf{5}}_{1,3}, ~\overline{\mathbf{5}}_{1,5}, ~\overline{\mathbf{5}}_{3,4}, \\[4pt]
~T_{1,4}, ~\overline{T}_{1,4}, ~H_{4,5}, ~\overbar{H}_{4,5},~3\,S_{1,3},
~S_{1,4},~S_{4,1},~S_{1,5}, ~4\, S_{2,4}, ~S_{2,5}, ~3\,
S_{3,2}, ~S_{3,4}, ~3\, S_{3,5}  \; ,
\end{gathered}
\label{SMAb}
\end{equation}
where we have denoted the singlet fields by $S_{a, b}$ and
we have used the compressed $SU(5)$--notation, where appropriate. The spectrum contains, apart from the MSSM multiplets, a vector-like pair $T$ -- $\overline{T}$ of exotic quarks and a number of singlet matter fields, which correspond to bundle moduli. These singlet fields can be given VEVs, which corresponds to deforming the bundle away from the split locus and into a non-abelian bundle. However, not all deformations of the bundle lead to supersymmetric vacua. In fact, for our example, the following terms
\beq\label{Wsing}
W_{\text{sing}}~ \sim~ S_{1,4}^p S_{4,1}^p~,
\eeq
where $p\geq 2$, are allowed by the $U(1)$ symmetries.
These operators are the only possible contributions to the singlet superpotential, $W_{\rm sing}$.
If indeed present, they obstruct switching on VEVs $\langle S_{1,4}\rangle$ and
$\langle S_{4,1}\rangle$ simultaneously and hence, we
require that either $\langle S_{1,4}\rangle = 0$ or $\langle S_{4,1}\rangle = 0$.
With this assumption, the D-term equations can be satisfied for generic (small) VEVs of the
remaining singlet fields, indicating the existence of supersymmetric vacua near the split
locus $t_1=t_2=t_3=t_4=t_5$ in K\"ahler moduli space.

The superpotential is further constrained. At the abelian locus, the coupling $H\overbar H$ must
be absent, as indicated by the cohomology computations. However, the superpotential
coupling $\overbar H_{4,5} L_{1, 5} S_{4,1}$ is allowed by the $U(1)$ symmetries.
We assume that $\langle S_{4,1}\rangle = 0$ in order to avoid generating a large Higgs mass from this term.

For the discussion of the QCD axion, the relevant superpotential couplings allowed by the $U(1)$ symmetries are
\beq
W\supset \overline{ T}_{1,4} \,d_{3,4}\, S_{1,3} ~.
\label{superpot}
\eeq
For a non-zero $S_{1,3}$ VEV, this coupling removes the $d_{3,4}$ -- $\overline{ T}_{1,4}$ pair from the massless spectrum and, hence, these fields play the role of the exotic quark fields $\cQ$ and $\widetilde\cQ$ from our general set-up. (The ``missing" $d$-type quark is replaced by $T_{1,4}$ which carries the same standard model quantum numbers.) Altogether, this provides a realisation of the axion mechanism discussed in the previous sections. If $\langle S_{1,3}\rangle$ can be stabilised at a small value, $10^{-7}\, \lsim\, \langle S_{1,3}\rangle\,\lsim\,10^{-4}$ in GUT units, the axion coupling parameter will be in the phenomenologically allowed range.

\section{Summary and conclusions}
In this note, we have shown that a KSVZ axion with a decay constant in the phenomenologically required range can be realised in the context of heterotic Calabi-Yau compactifications with split bundles. At the split locus, the low-energy symmetry is enhanced by one or several $U(1)$ factors. Their associated FI terms vanish at the split locus and can assume arbitrarily small values close to it. Hence, solving the D-term equations in the vicinity of the split locus leads to a small VEV for a matter field singlet. Provided this singlet is coupled to a pair of exotic quarks, its phase becomes an axion with a decay constant set by the size of the FI term. We have presented an explicit line bundle standard model where all the required ingredients are present.

Hence, a phenomenologically viable axion scale can be obtained provided the moduli are dialled to the right values close to the split locus. This shows that there is no in-principle obstruction to implementing the axion solution to the strong CP problem in the context of the heterotic string. However, in this note we have not attempted to {\it explain} the axion scale, that is, to stabilise the moduli in the required region of moduli space. While it is not implausible that moduli are stabilised in the vicinity of a locus with enhanced symmetry, implementing this explicitly remains the subject of future work.

In the present note, we have presented one explicit example. It is worth noting that the database \cite{lbdatabase} contains a large number of potentially interesting models, thus opening up a large area for exploring axion physics in heterotic string theory.

\section*{Acknowledgements}
We would like to thank Kiwoon Choi for helpful discussions.
The work of  EIB~is supported by the ARC Future Fellowship FT120100466. AL~is partially supported by the EPSRC network grant EP/l02784X/1 and by the STFC grant~ST/L000474/1. EIB and AC~would like to thank the Theoretical Physics Department at Oxford University for hospitality during part of the preparation of this paper.


\newpage

\begin{thebibliography}{10}

\bibitem{Peccei:1977hh}
R.~Peccei and H.~R. Quinn, ``{CP Conservation in the Presence of Instantons},''
  {\em Phys.Rev.Lett.} {\bf 38} (1977)
1440--1443.

\bibitem{Kim:1986ax}
J.~E. Kim, ``{Light Pseudoscalars, Particle Physics and Cosmology},'' {\em
  Phys.Rept.} {\bf 150} (1987)
1--177.

\bibitem{Svrcek:2006yi}
P.~Svrcek and E.~Witten, ``{Axions In String Theory},'' {\em JHEP} {\bf 0606}
  (2006) 051,
\href{http://arXiv.org/abs/hep-th/0605206}{{\tt hep-th/0605206}}.

\bibitem{Conlon:2006tq}
J.~P. Conlon, ``{The QCD axion and moduli stabilisation},'' {\em JHEP} {\bf
  0605} (2006) 078,
\href{http://arXiv.org/abs/hep-th/0602233}{{\tt hep-th/0602233}}.

\bibitem{Dasgupta:2008hb}
K.~Dasgupta, H.~Firouzjahi, and R.~Gwyn, ``{On The Warped Heterotic Axion},''
  {\em JHEP} {\bf 0806} (2008) 056,
\href{http://arXiv.org/abs/0803.3828}{{\tt 0803.3828}}.

\bibitem{Honecker:2013mya}
G.~Honecker and W.~Staessens, ``{On axionic dark matter in Type IIA string
  theory},'' {\em Fortsch.Phys.} {\bf 62} (2014) 115--151,
\href{http://arXiv.org/abs/1312.4517}{{\tt 1312.4517}}.

\bibitem{Choi:2014uaa}
K.~Choi, K.~S. Jeong, and M.-S. Seo, ``{String theoretic QCD axions in the
  light of PLANCK and BICEP2},'' {\em JHEP} {\bf 1407} (2014) 092,
\href{http://arXiv.org/abs/1404.3880}{{\tt 1404.3880}}.

\bibitem{Anderson:2009nt}
L.~B. Anderson, J.~Gray, A.~Lukas, and B.~Ovrut, ``{Stability Walls in
  Heterotic Theories},'' {\em JHEP} {\bf 0909} (2009) 026,
\href{http://arXiv.org/abs/0905.1748}{{\tt 0905.1748}}.

\bibitem{Anderson:2011ns}
L.~B. Anderson, J.~Gray, A.~Lukas, and E.~Palti, ``{Two Hundred Heterotic
  Standard Models on Smooth Calabi-Yau Threefolds},'' {\em Phys.Rev.} {\bf D84}
  (2011) 106005,
\href{http://arXiv.org/abs/1106.4804}{{\tt 1106.4804}}.

\bibitem{Anderson:2012yf}
L.~B. Anderson, J.~Gray, A.~Lukas, and E.~Palti, ``{Heterotic Line Bundle
  Standard Models},'' {\em JHEP} {\bf 1206} (2012) 113,
\href{http://arXiv.org/abs/1202.1757}{{\tt 1202.1757}}.

\bibitem{Anderson:2013xka}
L.~B. Anderson, A.~Constantin, J.~Gray, A.~Lukas, and E.~Palti, ``{A
  Comprehensive Scan for Heterotic SU(5) GUT models},'' {\em JHEP} {\bf 1401}
  (2014) 047,
\href{http://arXiv.org/abs/1307.4787}{{\tt 1307.4787}}.

\bibitem{He:2013ofa}
Y.-H. He, S.-J. Lee, A.~Lukas, and C.~Sun, ``{Heterotic Model Building: 16
  Special Manifolds},''
\href{http://arXiv.org/abs/1309.0223}{{\tt 1309.0223}}.

\bibitem{Anderson:2014hia}
L.~B. Anderson, A.~Constantin, S.-J. Lee, and A.~Lukas, ``{Hypercharge Flux in
  Heterotic Compactifications},''
\href{http://arXiv.org/abs/1411.0034}{{\tt 1411.0034}}.

\bibitem{Buchbinder:2013dna}
E.~I. Buchbinder, A.~Constantin, and A.~Lukas, ``{The Moduli Space of Heterotic
  Line Bundle Models: a Case Study for the Tetra-Quadric},'' {\em JHEP} {\bf
  1403} (2014) 025,
\href{http://arXiv.org/abs/1311.1941}{{\tt 1311.1941}}.

\bibitem{Buchbinder:2014qda}
E.~I. Buchbinder, A.~Constantin, and A.~Lukas, ``{A heterotic standard model
  with $B - L$ symmetry and a stable proton},'' {\em JHEP} {\bf 1406} (2014)
  100,
\href{http://arXiv.org/abs/1404.2767}{{\tt 1404.2767}}.

\bibitem{Buchbinder:2014sya}
E.~I. Buchbinder, A.~Constantin, and A.~Lukas, ``{Non-generic Couplings in
  Supersymmetric Standard Models},''
\href{http://arXiv.org/abs/1409.2412}{{\tt 1409.2412}}.

\bibitem{Kim:1979if}
J.~E. Kim, ``{Weak Interaction Singlet and Strong CP Invariance},'' {\em
  Phys.Rev.Lett.} {\bf 43} (1979)
103.

\bibitem{Shifman:1979if}
M.~A. Shifman, A.~Vainshtein, and V.~I. Zakharov, ``{Can Confinement Ensure
  Natural CP Invariance of Strong Interactions?},'' {\em Nucl.Phys.} {\bf B166}
  (1980)
493.

\bibitem{Blumenhagen:2005ga}
R.~Blumenhagen, G.~Honecker, and T.~Weigand, ``{Loop-corrected
  compactifications of the heterotic string with line bundles},'' {\em JHEP}
  {\bf 0506} (2005) 020,
\href{http://arXiv.org/abs/hep-th/0504232}{{\tt hep-th/0504232}}.

\bibitem{Kaplunovsky:1994fg}
V.~Kaplunovsky and J.~Louis, ``{Field dependent gauge couplings in locally
  supersymmetric effective quantum field theories},'' {\em Nucl.Phys.} {\bf
  B422} (1994) 57--124,
\href{http://arXiv.org/abs/hep-th/9402005}{{\tt hep-th/9402005}}.

\bibitem{Horava:1995qa}
P.~Horava and E.~Witten, ``{Heterotic and type I string dynamics from
  eleven-dimensions},'' {\em Nucl.Phys.} {\bf B460} (1996) 506--524,
\href{http://arXiv.org/abs/hep-th/9510209}{{\tt hep-th/9510209}}.

\bibitem{Horava:1996ma}
P.~Horava and E.~Witten, ``{Eleven-dimensional supergravity on a manifold with
  boundary},'' {\em Nucl.Phys.} {\bf B475} (1996) 94--114,
\href{http://arXiv.org/abs/hep-th/9603142}{{\tt hep-th/9603142}}.

\bibitem{Lukas:1997fg}
A.~Lukas, B.~A. Ovrut, and D.~Waldram, ``{On the four-dimensional effective
  action of strongly coupled heterotic string theory},'' {\em Nucl.Phys.} {\bf
  B532} (1998) 43--82,
\href{http://arXiv.org/abs/hep-th/9710208}{{\tt hep-th/9710208}}.

\bibitem{Lukas:1998hk}
A.~Lukas, B.~A. Ovrut, and D.~Waldram, ``{Nonstandard embedding and five-branes
  in heterotic M theory},'' {\em Phys.Rev.} {\bf D59} (1999) 106005,
\href{http://arXiv.org/abs/hep-th/9808101}{{\tt hep-th/9808101}}.

\bibitem{lbdatabase}
The database of heterotic line bundle standard models obtained in
  \cite{Anderson:2011ns} and \cite{Anderson:2013xka} can be accessed at {\tt
  http://www-thphys.physics.ox.ac.uk/projects/CalabiYau/linebundlemodels/}.

\bibitem{Braun:2010vc}
V.~Braun, ``{On Free Quotients of Complete Intersection Calabi-Yau
  Manifolds},'' {\em JHEP} {\bf 1104} (2011) 005,
\href{http://arXiv.org/abs/1003.3235}{{\tt 1003.3235}}.

\bibitem{Candelas:2008wb}
P.~Candelas and R.~Davies, ``{New Calabi-Yau Manifolds with Small Hodge
  Numbers},'' {\em Fortsch.Phys.} {\bf 58} (2010) 383--466,
\href{http://arXiv.org/abs/0809.4681}{{\tt 0809.4681}}.

\bibitem{Candelas:2010ve}
P.~Candelas and A.~Constantin, ``{Completing the Web of $Z_3$ - Quotients of
  Complete Intersection Calabi-Yau Manifolds},'' {\em Fortsch.Phys.} {\bf 60}
  (2012) 345--369,
\href{http://arXiv.org/abs/1010.1878}{{\tt 1010.1878}}.

\end{thebibliography}

\providecommand{\href}[2]{#2}\begingroup\raggedright\endgroup

\end{document}